\documentclass[a4paper]{article}
\usepackage{graphicx}
\usepackage{rsc}         
\usepackage[pdftex]{geometry}
\usepackage{natmove}
\usepackage{subfig}

\usepackage{xcolor}
\usepackage{amsmath}

\usepackage{pdfpages}

\begin{document}
\title{Solvent-induced ion clusters generate long-ranged double-layer forces at high ionic strengths}
\author{ {David Ribar*, Clifford E. Woodward** and Jan Forsman*}
\\  
$^*$Computational Chemistry, Lund University
\\
 P.O.Box 124, S-221 00 Lund, Sweden
\\
$^{**}$School of Physical, Environmental and Mathematical Sciences
\\
University College,  University of New South Wales, ADFA
\\
Canberra ACT 2600, Australia   
\\
}

\maketitle
\abstract{
    Recent experimental
    results by the Surface Force Apparatus (SFA) have identified a dramatic deviation from
    previously established theories of simple electrolytes.  This deviation, referred to as anomalous
    underscreening, suggests that the range of electrostatic interactions increase upon
    a further addition of salt, beyond some threshold concentration (usually about 1M).
    In this theoretical work, we explore an extension of the
    Restricted Primitive Model (RPM) wherein a short-ranged pair potential of mean force (sPMF) is added
    to the usual Coulombic interactions so as to mimic
    changes of the hydration as two ions approach one another. 
    The strength of this potential is adjusted so that the modified RPM
    saturates at a realistic concentration level (within a range  4-7M,  typical to aqueous 1:1 salts).
    We utilise grand canonical simulations to establish surface forces predicted by the model and
    compare them directly with SFA data. 
    { We explore different sPMF models, which in all cases display significant clustering at concentrations above about 1M.
    In these models, we find significant double-layer repulsion at separations that significantly exceed those
    expected from standard RPM predictions. We do not, however, observe an
    {\em increase} of the screening length with salt concentration, but rather
    that this screening length seemingly saturates at a (rather high) value.}
    The simulated long-ranged interactions are shown to correlate with ion cluster
    formation, implicating the important role of accompanying {\em cluster-cluster} interactions. In particular, steric
    interactions between clusters (manifested in density-density correlations) are quite relevant in these systems.
}

\section{Introduction}

The theoretical study of interactions between large charged particles is important in almost all areas
of colloid science \cite{Israelachvili:91,Evans:94,Holm:01}. For some time there had been a general consensus 
that the behaviour of charged particles in aqueous solutions of monovalent (1:1) electrolytes 
could be described by so-called primitive models (PM), at least qualitatively.
In these models the aqueous solvent is implicitly treated, only asserting itself via a uniform relative dielectric
constant, $\varepsilon_r$.
The ionic species, on the other hand, are explicitly accounted for as charged hard spheres. In a mean-field treatment
that ignores ion-ion correlations, this description can be even further simplified by neglecting the hard core interactions 
between the background salt ions.  Meanwhile, colloidal particles can be treated as charged surfaces. 
Assuming additive van der Waals forces (of quantum origin)  between the surfaces results in the famous 
DLVO theory \cite{Derjaguin:41,Verwey:48}, which has proven
very useful for the study of the stability of colloidal suspensions.

One generally finds that mean-field (MF) treatments of the PM predict that, at low
electrolyte concentrations, the dominant non-quantum interaction
between charged surfaces is electrostatic in origin and is 
exponential in form, with a range largely dictated by the Debye screening length, $\lambda_D$. 
This interaction arises from the direct forces between the charged surfaces that are screened by 
a counter-charge density sourced from the background salt.  The Debye length measures the typical spatial
length over which non-electroneutral fluctuations occur in the electrolyte solution.  
Structural variations in the overall density (rather than charge density) 
are also induced by the surfaces, but the forces that arise
from this effect are usually negligible at large surface separations. \cite{Attard_1993,LeotedeCarvalho1994}
The Debye length decreases with electrolyte concentration, $c$, according to, $\lambda_D \sim 1/\sqrt{c}$.
Corrections to MF theories require the inclusion of direct (short-ranged) ion correlations, which
lead to minor modifications of the Debye length.  For example, taking into account hard core
interactions leads to an excluded volume corrected 'effective' value $\lambda_{eff}$, which is smaller
than $\lambda_D$ \cite{Attard_1993, Forsman:24a}.
The qualitative correctness of MF theories using the PM, and 
the role played by the Debye length, is reasonably well-supported  
at low electrolyte concentrations by  experiments, as well as 
simulations. \cite{Israelachvili:91}  

MF theories break down at high concentration where the Debye length becomes small.
In this regime, ion correlation corrections {\em must} be included to provide even 
a qualitative description of the PM.   
Such corrections predict that the surface interactions 
of electrostatic origin will switch from exponential to oscillatory, as the Debye length 
approaches the diameter scale of the salt particles.  This is the so-called Kirkwood transition. \cite{Kirkwood1936}  
This transition has been confirmed by simulations 
of the PM and is due to the monotonically  decreasing (essentially exponential) ionic double-layer 
becoming unstable to the  formation of charged layers by the electrolyte. \cite{Torrie:80}
In this regime, the  non-electrostatic structural forces may compete with the 
electrostatic interactions for relative importance as both are oscillatory and of 
comparable range. \cite{Attard_1993,LeotedeCarvalho1994}   However, recent experimental measurements
using the Surface Force Apparatus (SFA) 
have challenged this result \cite{Gebbie:13,Gebbie:15,Smith:2016,Elliot:24}.
That is, instead of transitioning to an oscillatory function, the asymptotic 
interaction between charged surfaces remains exponential, but with a screening length that suddenly begins 
increasing (rather than decreasing) with concentration,
eventually becoming significantly larger than the nominal Debye length.
Surprisingly, this occurs even in aqueous 1:1 salt solutions, where it
was generally believed that the PM should be reasonably reliable. 
This phenomenon has been described as {\em anomalous underscreening}.
One consequence of this is that suspensions of like-charged colloids may remain stable at high ionic strength 
due to significant repulsive interactions at colloidal separations where the attractive van der Waals interactions
remain small.   This is counter to the predictions of traditional DLVO theory.  
Despite considerable theoretical efforts, \cite{coupette:18,Rotenberg:18, Kjellander2018,Coles:20,Zeman:20,Cats:21,Hartel:23} 
there is at present no consensus as to the physical origin of interactions at such large inter-particle distances.  
{One can find independent (non-SFA) experimental evidence for anomalous
   underscreening in literature \cite{Gaddam:19,Yuan:22, Reinertsen2024, Robertson2023}. It should also be
   noted that there is at least one recent  experimental work that 
   {\em does not} support this phenomenon\cite{Kumar:22}. Kumar {\em et al.} used an atomic force microscope to measure interactions between
   charged surfaces (mainly silica-silica) in aqueous salt solutions. They did not find any sign of anomalous underscreening, i.e.
   at high ionic strengths, the measured surface forces were quite short-ranged.}

The  formation of ionic clusters of low overall charge has been proposed as
a possible mechanism for the putative increase in the Debye length observed in the 
underscreening concentration regime.  Such a model implicitly assumes that the electrostatic interactions between the
surfaces remains dominant and the formation of clusters 
will effectively reduce the concentration of screening charges \cite{Gebbie:15,Ma:15}.
For this to be a feasible explanation, clustering needs to be sufficient  
to reduce the free charge concentration by a factor of up to $\sim 10^4$.
Recent simulations of the {\em Restricted} Primitive Model (RPM), wherein ions have equal sizes, have investigated
the degree of clustering that occurs in a 1:1 electrolyte \cite{Hartel:23}. That work also took into 
account dielectric saturation by allowing the relative dielectric constant of the solvent,  $\varepsilon_r$,  
to become concentration-dependent.  In particular $\varepsilon_r$  decreased with $c$
 to account for the diminished dielectric response of rotationally constrained
water molecules in the ionic hydration shells. 
\cite{hasted_dielectric_2004, deSouza2022, Conway83, Bonthuis2012, DanielewiczFerchmin2013, adar_dielectric_2018, Underwood2022}  
It was seen that, while increased clustering was evident at high
concentrations, and lower $\varepsilon_r$, it was not of a sufficient magnitude to 
explain anomalous underscreening. \cite{Hartel:23}  That is, the RPM 
(even accounting for dielectric saturation) would 
predict that forces between like-charged surfaces would be essentially completely screened 
by salt ions at the large distances where experiments still find a significant repulsion.  

This failure of the RPM led to a recent work \cite{Ribar:24}
in which we proposed a {\em localised} modification of the RPM by addition of 
an additional short-ranged potential of mean force (sPMF) between ions.
The inclusion of the short-ranged potential induced significant ion clustering (as 
confirmed via simulations). In the present article, we study the consequences of 
the formed clusters on the interactions between two charged surfaces. These interactions are directly compared
  with experimental SFA data, and it is demonstrated that cluster formation indeed leads to remarkably
  long-ranged surface forces. We provide analyses of the separate contributions to the correlations
  in these systems in order to uncover the dominant physical mechanism leading to these long-ranged forces.

\section{Model and methods}
A weakness of the RPM is that it does not (with an aqueous solvent) saturate at reasonable concentrations.
In this work, we explore the hypothesis that what the RPM is lacking is a short-ranged potential
of mean force (sPMF), adjusted to induce
solution instability at a concentration that is close to typical saturation values (4-7M) for simple salts. While we do not
have a detailed knowledge of this potential of mean force, it is plausibly a result of 
water restructuring as the ionic environment becomes crowded.  
In previous work, \cite{Ribar:24} we took
an approach whereby we assigned the sPMF a range of about one water diameter (3 {\AA}). Since our simulations are
based on Coulomb interactions, a {\em convenient} option was to increase the strength of the Coulomb interaction in this separation
range.  This is achieved by varying the dielectric constant using linear extrapolation from the bulk down
to a smaller contact value.  
{ Of course, this choice of additional short-ranged interaction is not unique and 
in this paper we will consider different functional forms of the sPMF, including
the model we used earlier.\cite{Ribar:24}  Due to its relatively narrow range, we will dub that model the ``narrow'' sPMF. 

\subsection{The ``narrow'' sPMF, $\phi_n$}

As described above, the explicit form for this sPMF uses the following linear ramp function for the spatial variation of 
the dielectric constant, $\varepsilon_r$, within its range,
\begin{equation}
	\varepsilon_r(r) = \left\{
	\begin{array}{ll}
		\varepsilon_c ; & r \leq d \\
		\varepsilon_c + (\varepsilon_b - \varepsilon_c) \frac{r - d}{\Delta}; & d < r \leq d + \Delta  \\
		\varepsilon_b; & r > d + \Delta  \\
	\end{array}
	\right..
	\label{eq:vareps_def}
\end{equation}
Here, $d$, is the hard-sphere diameter of the ions, while $\Delta$, is a measure of the
thickness of a hydration shell and chosen to be $\Delta=3$ Å, the diameter of a water molecule.
The parameters, $\varepsilon_c$ and $\varepsilon_b$ are the \textit{contact} and \textit{bulk} 
values of the dielectric factor, respectively. At room temperature, $\varepsilon_b = 78.3$
and we have chosen, $\varepsilon_c < \varepsilon_b$.  It could be argued that 
this inequality reflects the exclusion of water solvent as ions approach each other, together with the strengthening of the 
electric field leading to dielectric saturation.   However, as we discuss later, these effects are really many-body
in character and can only qualitatively be reproduced via an sPMF, which is a 2-body interaction.  More pragmatically, 
this approach can be viewed as merely a convenient way to introduce a solvent-induced PMF
at short range which promotes solution saturation and does not mean that the dielectric factor {\em actually} varies in this way.
Indeed, we would argue that since the effects we attempt to incorporate are effective at the molecular scale, they 
are not describable by continuum electrostatics. 

With this choice for the dielectric factor,  the sPMF between ions $i$ and $j$ of valency $|Z_i| = |Z_j| = 1$ ($\phi_n^{ij}$) can be written as:
\begin{equation}
  \beta\phi_n^{ij}(r) =
  \left\{
	\begin{array}{ll}
		0 ; & r > d+\Delta \\
		\frac{Z_i Z_j}{r}(l_B(r)-l_B(bulk)); & d < r < d+\Delta   \\
	\end{array}
	\right.
	\label{eq:phin}
\end{equation}
where $l_B(r) = \beta e_0^2 /(4\pi\varepsilon_0\varepsilon_r(r))$ and $l_B(bulk) = \beta e_0^2 /(4\pi\varepsilon_0\varepsilon_b)$.
It should be noted that the nature of $\phi_n^{ij}$ is given by the valency of the interacting
species, i.e. it is always attractive between unlike charges, and repulsive between charges of equal sign.

For a given choice of $d$, we can tune $\varepsilon_c$ to a value where the solution phase separates
at a concentration typical of the saturation limit for an aqueous monovalent electrolyte solution
(about 4-7 M, at room temperature).  In practice, we chose  $d=3$ {\AA} and $\varepsilon_c=23$ where the
solution is stable at 3.45 M, but will phase separate at a marginally smaller value for $\varepsilon_c$.  Our rationale
for choosing these values in given in \cite{Ribar:24}.  We also consider $d=4$ {\AA}, in which
case $\varepsilon_c=20$ will give a similar stability.  
We will use this as the main model of investigation in this paper, including looking at the effect 
of  salt concentration on the interaction between charged surfaces. 
  
  We will also perform a more limited study on a so-called  "wide" version of an sPMF, 
  where we will also explore other different combinations of attractions and repulsions
  between ionic species.  The details of this interaction are given below.

\subsection{The ``wide'' sPMF, $\phi_w$ }
Here, we adopt a sPMF with the same cut-off ($\Delta$) as $\phi_n^{ij}$, but with a
broader range over which it maintains a substantial strength. Specifically, we define it as,
\begin{equation}
  \beta\phi_w^{ij}(r) =
  \left\{
	\begin{array}{ll}
		0 ; & r > d+\Delta \\
                A_{ij}\left(\left(\frac{r-d}{d}\right)^4-1\right); & d < r < d+\Delta   \\
	\end{array}
	\right.
	\label{eq:phiw}
\end{equation}
One important difference, compared with the narrow version, is that the sign
of $A_{ij}$ can be arbitrarily assigned. This leads to various sPMF:s
with a qualitatively different impact.  The following nomenclature describes the choice of parameter values 
we used in this study. For instance,
$\phi_w(a,r)$ indicates that $A_{+-}$ is positive between
unlike charges (attractive sPMF), but that $A_{+-} = -A_{--} = -A_{++} = 2.5$.
The corresponding notation for an overall attractive sPMF
is $\phi_w(a,a)$, for which $A_{++} = A_{--} = A_{+-} = 0.5$.
Finally, $\phi_w(a,0)$ signifies an attractive sPMF
between unlike charges, but with no impact on the interaction between ions
of the same sign, i.e. $A_{++} = A_{--} = 0$, and $A_{+-}=1.0$.
In all cases, the amplitudes used were regulated to be strong
enough to ``almost'' induce phase separation in a bulk solution at 3.45 M.
We have only investigated $\phi_w$ systems in which $d=3$ {\AA}.

The overall qualitative difference between $\phi_n(r)$ and $\phi_w(r)$ is illustrated in
Figure \ref{fig:phicomp}.
\begin{figure}
  \centering
    	\includegraphics[width=8.7cm]{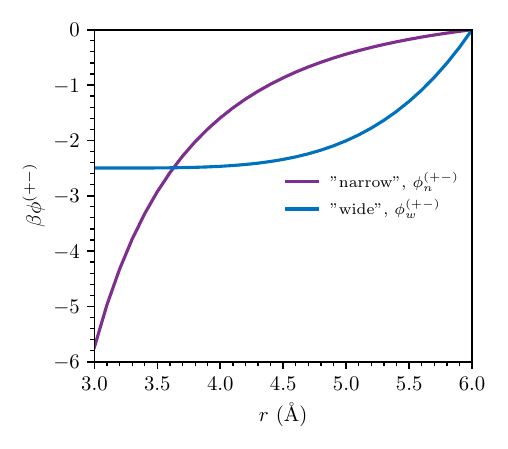}  
	\caption{{A comparison between $\phi_n(r)$ and $\phi_w(r)$. The illustration
          is made for the $+-$ sPMF, which is always attractive, and for $d=3$ {\AA}.
          Moreover, for the narrow sPMF, we have set $\varepsilon_c=23$, whereas
          the wide sPMF is based on an amplitude $A_{+-}=2.5$.
           }}
	\label{fig:phicomp}
\end{figure}
The {\em total} potential of mean force between ions $i$ and $j$, $\phi_{ij}(r)$ (separated a distance $r$)
can thus be written as:
\begin{equation}
		\beta\phi_{ij}(r) = \left\{
	\begin{array}{ll}
		\infty ; & r \leq d \\
		l_B(bulk)\frac{Z_i Z_j}{r} + \beta\phi_\alpha^{ij}(r); & r > d  \\
	\end{array}
	\right.
	\label{eq:phitot}
\end{equation}
where $\alpha=$ $w$ or $n$, depending on the choice of sPMF.
} Importantly, for $r \ge d + \Delta$, the pair potential is equivalent to the RPM pair potential, corroborating the
local nature of our modification.


{ \subsection{Surface force simulation details}}
Our simulation system is comprised two parallel flat and uniformly charged surfaces, separated at a distance $H$, and extending
infinitely along the $x$ and $y$ directions. We imagine these surfaces to be immersed in a
salt solution.  The bulk region
external to the space between the surfaces is characterised by the electro-chemical
potential and the bulk osmotic pressure, $p_b$. A grand canonical
simulation scheme ensured that the simulated fluid, contained between the surfaces, was
in equilibrium with the bulk and that the surface charges were properly
neutralised. \cite{Stenberg:2021a}.
{The surface charge density was chosen to be $\sigma_s=1/70\approx -0.014336 \; e/${\AA}$^2$,
which is a value typical to mica \cite{Crothers:21}, i.e. the surfaces used in SFA experiments.}
All simulations were performed at 298 K.
Periodic boundary conditions were applied along the  the $(x,y)$ directions, and
the ``charged sheet'' method \cite{Torrie:80} was adopted to
account for long-ranged interactions. Cluster moves were implemented, leading to
crucial improvements of the statistical performance.
The internal (osmotic) pressure, $p$, acting transverse to the walls was evaluated from the
average normal force across the walls, separated by the expected distance
between Gibbs dividing (solid-fluid) surfaces, $h \equiv H - d$. Subtracting $p_b$, we obtain the net pressure, $p_{net} = p - p_b$.
Integrating this pressure, and utilising the Derjaguin Approximation 
{
\begin{equation}
F(z)/R = - 2\pi \int_\infty^zp_{net}(z) \; dz
\end{equation}
}
allows us to construct force per radius curves, $F/R$, facilitating direct comparisons with SFA data ($R$ is the 
radius of the curved surface used in the experimental setup).
To make comparisons with the SFA data, the simulated interaction curves were matched 
to the experimental results for a 2 M NaCl solution at a separation of 57 {\AA}.

{ The non-electrostatic ion-wall interaction was modelled by an analytic and purely repulsive
  wall potential. We explored two different choices, decaying as $\delta^{-4}$
and $\delta^{-6}$, respectively, where $\delta$ is the transverse distance
between the ion and the wall. The first model was adopted
when investigating the effect of salt concentration, using $\phi_n$ as the sPMF.
One might argue that this softer choice amounts to a coarse-grained
description of molecularly rough surfaces. In another part of this study, which
explores the various choices for the sPMF in concentrated solutions (1.6-2 M), we
chose the steeper wall, decaying as $\delta^{-6}$. This represents a better
model for a molecularly smooth surface, such as mica. 
Details are provided in the Electronic Supporting Information, ESI, including a comparison between
results obtained with either wall, for a given system. The
ESI also contains a more thorough description of our model and simulation methods.}

\section{Results and discussion}
{
Our results comprise two parts. In the first part, we
evaluate how surface forces respond to salt concentration changes, using the
``soft' wall ($\delta^{-4}$) description (see ESI), and $\phi_n$ as the sPMF.
The second part compares surface interactions at a
high salt concentration, using various choices for the sPMF.
In that part, we have employed the ``steep'' wall ($\delta^{-6}$) description, which
arguably is a better choice for mica surfaces. On the other hand, as we discuss in the ESI, the
difference in wall softness essentially amounts to a small shift (about 3{\AA}) of the surface separation.} 

\subsection{Salt concentration effects, using $\phi_n$}
{ In previous work we performed simulations  using the "narrow" $\phi_n$ potential as the sPMF 
in addition to the standard RPM interaction for bulk electrolyte solutions \cite{Ribar:24}.  We will denote this as the sRPM($\phi_n$) model}  
The findings important to the current investigation can be summarised 
as follows: 
\begin{enumerate}
	
	\item { At low electrolyte concentration, charge-charge correlations displayed behaviours typical
		of classic MF theory, and the Debye length is the relevant length-scale.}
	\item{ At higher concentration (beyond 1 M) there was
		evidence of strong cluster formation, much larger than predicted by the RPM.  Charge-charge correlations
		became short-ranged and oscillatory (the Kirkwood transition), while
		density-density correlations grew to become more dominant, displaying
		a monotonic decay, with a length-scale much longer than the Debye length.}
	
\end{enumerate} 
\subsubsection{Simulated surface forces}
\begin{figure}[h!]
  \centering
  \includegraphics[scale=1]{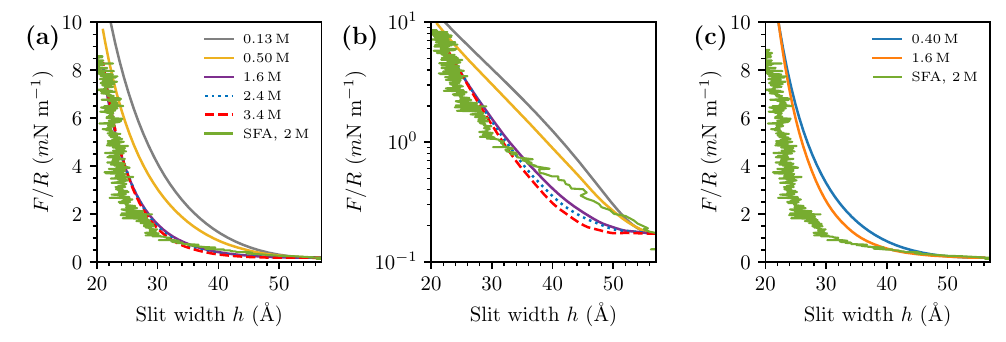}
    \caption{Simulated prediction of $F/R$, with $d=3$ {\AA} and $d=4$ {\AA}, at various salt concentrations, 
      compared with SFA data with a 2 M $NaCl_{(aq)}$ solution, obtained
      from prof. Susan Perkin (data previously published in Ref. \cite{Smith:2016}). Simulated curves have been shifted
      to roughly match the SFA data, at the largest investigated separation.
      (a) $d=3$ {\AA} (linear scale).
      (b) $d=3$ {\AA} (log scale). (c) $d=4$ {\AA} (linear scale.}
	\label{fig:d3}
\end{figure}

The presence of large clusters in the sRPM($\phi_n$) model prompted us to consider
their consequence on the interaction between charged surfaces.
Specifically, we report here simulations of charged surfaces immersed 
in the  sRPM($\phi_n$) solution and compare our results with experimentally measured forces that show
anomalous underscreening.  
\begin{figure}[h!]
  \centering
    	\includegraphics[width = 8.7cm]{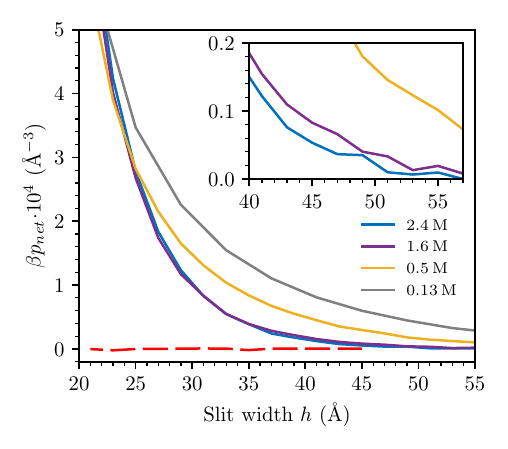}  
	\caption{Results from RPM($\phi_n$) grand canonical simulations of net pressures, at various bulk salt
          concentrations, with $d=3$ {\AA}, $\varepsilon_c=23$ and $\varepsilon_b=78.3$. An exception is
          the dashed line, which displays
          results using the standard RPM, with a uniform $\varepsilon_r=23$. The inset is a zoom-in at large
        separations.}
	\label{fig:P}
\end{figure}
Simulated net pressures acting on the charged surfaces for a range of different concentrations 
of the sRPM($\phi_n$) are shown in Figure \ref{fig:P}. We observe a significant repulsion for the simulated curves 
over a large range of surface-surface separations.
The range of the repulsion initially decreases as the ionic strength increases, 
but then appears to level out and become essentially concentration-independent 
above some threshold value ($<$ 1.6 M).  For comparison, we have 
also included results for the 1.6 M  RPM electrolyte in which the
relative dielectric constant was uniform and set to $\varepsilon_r=23$.
We found no detectable net pressure for the RPM electrolyte at any separation.  {A short-ranged
sPMF can induce 
clustering more efficiently than a global increase of the electrostatic coupling. This can be explained as follows. Consider
linear cluster of alternating charges: +-+-+-. A global increase of the Debye length will generate a significant repulsion
between next-nearest neighbours, but this is not so with 
a short-ranged sPMF that is attractive between ions of opposite charge but repulsive when the ions share the same valency.}
Similar results would also be obtained for an RPM model with $\varepsilon_r=78.3$.  That is, the sRPM($\phi_n$)
predicts surface forces of a much longer range than the traditional RPM, even if a broad 
a range of dielectric constants is used in the latter.
From our previous work, we found that the sRPM($\phi_n$) predicts a much higher
degree of clustering than the RPM, again despite the value of $\varepsilon_r$.  
It follows that the much longer-ranged forces observed 
in the sRPM($\phi_n$) in  Figure \ref{fig:P} is due to this increased cluster formation. 
The  $F/R$ predictions that follow from the net pressures are given in Figure \ref{fig:d3}.
These are compared with SFA data on $NaCl_{(aq)}$, at a concentration of 2 M. 
%
%
The simulation predictions at close to 2 M
agree reasonably well with experiments for separations less than about 40 {\AA}. 
To put this agreement into context, we note that the $F/R$  curves predicted by the RPM would be 
essentially zero if included in Figure \ref{fig:d3}.  Given the discussion above, this prompts us to assert that the large range 
of the forces seen in experiments is due to significant ionic clustering in the electrolyte.  
This assertion implies that the relevant mechanism is to be found in the behaviour
of the solution phase and is not, for example, a surface phenomenon.
Moreover, given our previous findings of the dominance of density-density correlation
in the high density regime (beyond 1 M), it is anticipated that the surface forces
are due to structural changes in the overall density, rather than the charge density, as the
surfaces approach.  That is, clusters will adsorb onto the surfaces, primarily 
due to charge and multipole interactions.  The adsorbed layers will begin to breakdown 
as the surfaces encroach each other, leading to a net repulsion with a length-scale  
of the typical cluster size.  We expect that the average cluster size should in principle grow with the
concentration of the electrolyte solution, perhaps explaining the apparent growth in screening length
observed in anomalous underscreening.  
This notwithstanding, the sRPM($\phi_n$) predictions are not completely consistent with
experiments.  Figure \ref{fig:d3} indicates that at large concentrations 
the range of the interactions  appears to reach a plateau where 
$F/R$ becomes concentration-independent.
This disagrees with the ``anomalous underscreening'' predictions by SFA, according to which
there is an increase of the decay length as the ionic strength increases in the high concentration regime.

For completeness, we include results obtained with a larger value for the hard-sphere diameter
of the ions, $d=4$ {\AA}. Recall also that we chose $\varepsilon_c=20$ at this diameter, as
described in detail in ref. \cite{Ribar:24}.
Comparisons with raw SFA data are given in Figure \ref{fig:d3}(c).
We note that the difference between the interactions obtained
at 0.40 M and 1.6 M is considerably smaller than we find by a similar comparison (0.50 M and 1.6 M), with $d=3$ {\AA}.

We also compare the simulated net pressures directly with a derivative
\begin{figure}[h!]
	\centering
	\includegraphics[scale=0.9]{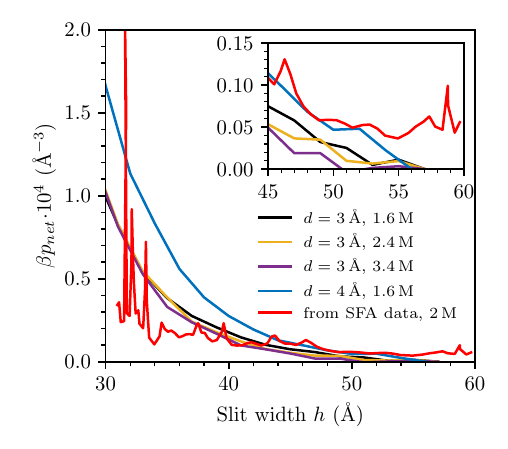}
	\caption{Simulated net pressures, and corresponding predictions from SFA data. The inset is a zoom-in at large
		separations.}
	\label{fig:der}
\end{figure}
of the experimentally observed free energies per unit area (taking into account an appropriate
scaling factor), Figure \ref{fig:der}. In the latter case, we needed to smooth out the raw data
by performing ``running averages'' prior to taking the derivative (via discrete difference).
The extremely long-ranged tail found by the SFA is perhaps even more
apparent in this presentation, and it appears to be an aspect of the experimental
data that the sRPM($\phi_n$) is unable to reproduce.

\subsubsection{Structural analyses}
Here we illustrate how the long range forces are related to
a slow density decay of the ionic solution, which results from cluster formation. We will 
only consider systems with $d=3$ {\AA}.

\begin{figure}[h!] 
    \centering
	\includegraphics[scale = 0.9]{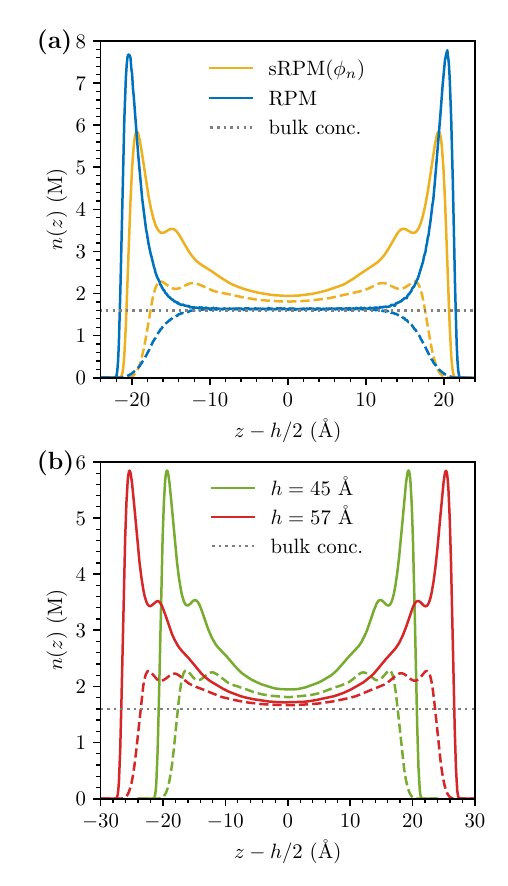}
        \caption{Ion concentration profiles along $z$. Solid lines depict cation density profiles, whereas the dashed lines show anion profiles.
            The dotted  line indicates the bulk concentration.  (a) Comparing concentration profiles at $h=48$ {\AA}, as obtained with
            the sRPM($\phi_n$) and the standard RPM.
    (b) Concentration profiles from sRPM($\phi_n$) simulations, at two different separations: $h=48$ {\AA} and $h=57$ {\AA}.
    }
    \label{fig:nz}
\end{figure}
  We start by analysing ion concentration profiles, $n(z)$, along the direction normal
  to the surfaces. 
In Figure \ref{fig:nz} (a), we note how the ion concentrations rapidly approach the bulk value in RPM simulations. However, using the
sRPM($\phi_n$), this approach is quite slow, and even at a surface separation of about 5 nm, both cation and anion mid plane concentrations
are {\em greater} the bulk value (recall that these results are from grand canonical simulations). This is because in the
many anions are members of  ion clusters, many of which are expected to be net positively charged and hence are adsorbed 
to the negative surface.  Thus they produce an anion concentration profile that approaches the mid plane from above.
By comparison, the RPM would produce an anionic profile approaching its mid plane value from from {\em below}. 

In Figure \ref{fig:nz} (b) we see how the mid plane values of the ion profiles gradually decreases towards the bulk
concentration, but we note that a significant difference persists even at quite large surface separations. 

\begin{figure}[h!]
	\centering
	\includegraphics[scale = 1]{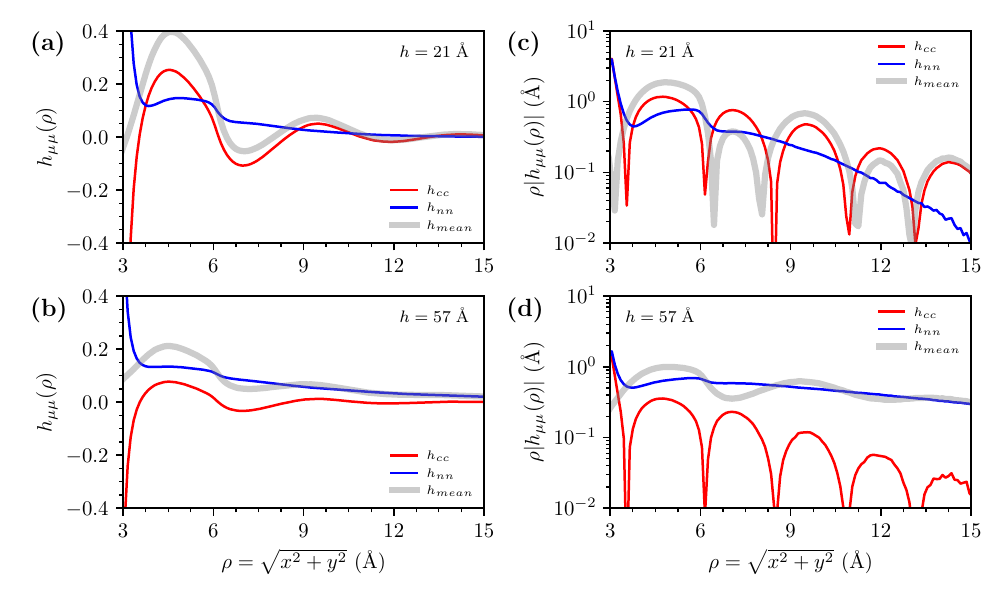}  
	\caption{Correlation function analysis in the slit, obtained for the $c \approx 3.4$ M system at
            the mid plane, for $h = 21$ Å and $h = 57$ Å respectively. Here, red is
			used for the charge-charge and blue for the density-density total correlation functions, while grey denotes the
			mean like-charge total correlation function. Note: $h_{mean} = h_{cc} + h_{nn}$.
            Graphs (a) and (b) present the $2D$ charge-charge, $h_{cc}$, and density-density, $h_{nn}$, total correlation
            functions on a linear scale. Graphs (c) and (d) provide the asymptotic analysis plots on a logarithmic scale.}
	\label{fig:correlations}
\end{figure}

  { We also investigated the} correlations parallel the surface, focusing on the region at the mid plane. 
  We computed lateral ($2D$) species resolved pair correlation functions (presented in detail in the ESI) for 
  low (0.13 M) and high (3.4 M) concentrations, for all slit widths under investigation. No evidence of a continuous solid phase
  was found in any of these systems indicating the absence of a frozen state. In order to further investigate the
  dominant correlations in our system, we
  computed the $2D$ charge-charge, $h_{cc}(\rho)$, and density-density, $h_{nn}(\rho)$, total correlation
  functions from the species resolved pair correlation functions \cite{LeotedeCarvalho1994}. Note that the total correlation functions are
  related to the pair distribution functions via $h(\rho) = g(\rho) - 1$. For brevity, the
  explicit $\rho = \sqrt{x^2+y^2}$ dependence is omitted.
\begin{align}
		h_{cc} &= \frac{1}{4}\left(h_{++} + h_{--} - 2h_{+-}\right) \\
		h_{nn} &= \frac{1}{4}\left(h_{++} + h_{--} + 2h_{+-}\right).
\end{align}
This allows us to essentially deconstruct the mean like-charge total correlation functions and observe the correlations dominating at
the asymptotic limit
via $h_{mean} = \frac{1}{2} (h_{++} + h_{--})= h_{nn} + h_{cc}$.   The asymptotic behaviour of correlation
  functions, as $r\to\infty$, is described by a Yukawa decay, or an oscillatory Yukawa
  decay \cite{LeotedeCarvalho1994}
	\begin{align}
		rh_{\mu\nu}(r) &\sim A^{(\mu\nu)}e^{ -\xi_0^{(\mu\nu)}r} \\
		rh_{\mu\nu}(r)& \sim A^{(\mu\nu)}e^{-\xi_0^{(\mu\nu)}r}\cos{\left[\xi_1^{(\mu\nu)}r-\Theta^{(\mu\nu)}\right]},
	\end{align}
	where $\mu\nu \in \{c, n, +, -\}$ denotes the corresponding correlation type $\mu\nu$ in a general sense.
	We define  $1/\xi_0^{(\mu\nu)}$ as the \textit{asymptotic} decay correlation
        length  (the effective screening length) and $ 2\pi/\xi_1^{(\mu\nu)}$ as the \textit{asymptotic} frequency correlation length, which
        describes the spatial range of electrolyte charged layer oscillations appearing after the
        Kirkwood transition. $A^{(\mu\nu)}$ is used for the amplitude and $\Theta^{(\mu\nu)}$ for the phase shift. Interested
        readers are referred to \cite{LeotedeCarvalho1994, Evans1994,Attard_1993} for detailed derivations. 
The results for the $c \approx$ 3.4 M system, obtained at the mid plane, is
presented on Figure \ref{fig:correlations}. Further details including all results from various slit widths under
investigation is presented in the ESI.
%
%
Figure \ref{fig:correlations}(a-b) presents the density-density and charge-charge total correlation functions on a linear scale
for two different surface
separations, $h = 21$ Å (top) and $h = 57$ Å (bottom). For the large surface separation we observe apparent monotonic decay of
the density-density correlations at long range and a damped oscillatory decay for the charge-charge correlations for both extreme surface separations.
In order to better observe the long-ranged asymptotic behaviour, we plot $\rho |h_{\mu\mu}({\rho})|, \mu\mu \in \{nn,cc\}$ using a log-linear
scale in Figure \ref{fig:correlations}(c-d). Here we observe two major changes in the correlation functions as the surface separation increases.
{Firstly at the smaller surface separations the charge-charge correlations display a higher amplitude compared to larger separations.}
Secondly, the density-density correlations decay faster at the shorter separation, with a much longer range decay at larger separations.
This is consistent with the cluster picture described above.  Namely at shorter separations, typical clusters will be smaller with highly
correlated charge-charge interactions, corresponding to more tightly packed ions.  While at larger separations, much larger
clusters can be accommodated
between the surfaces.  Such clusters will have a more loose charge arrangement.  It is precisely the exclusion
of larger clusters as the surface separation decreases which leads to the repulsion between the surfaces.
Indeed, this is the major conclusion of this paper.  This notwithstanding, our results  have by no means proven that 
substantial ion clusters do indeed exist in reality for such systems.  Our sRPM($\phi_n$) potential {\em approximates} an effect
due to the interaction between salt ions and the solvent, which should in principle be
modelled by many-body interactions in a solute-only approach. It should be noted that some
all-atomistic Molceular Dynamics simulations do suggest that ionic clustering can be quite
significant in the presence of explicit water models \cite{Degreve:99,Komori:23}. On the other hand,
such findings are quite dependent upon the potential model employed  \cite{Auffinger:07,Tong:23}.

{In order to complement the investigation of lateral correlation functions obtained at the mid plane, we present a comparison with
  bulk correlation functions on Figure \ref{fig:comp_correlations}, computed with the method described in our previous work \cite{Ribar:24}. We
  explicitly investigate three extreme cases: the smallest $h = 21$ Å surface separation, largest $h = 57$ Å, and a neutral wall $h = 77$ Å, by
  directly comparing with bulk correlation functions for a system of equal concentration. We plot all
  correlation functions on a linear scale on Figure \ref{fig:comp_correlations}(a). We can observe a minimal difference between
  the full ($h = 57$ Å) and dashed $(h = 77$ Å neutral walls) lines, insert on Figure \ref{fig:comp_correlations}(a).  All of the correlations in the
  slit have amplitudes lower than the bulk. Figures \ref{fig:comp_correlations}(b-d) demonstrate
  the similarity of the frequency correlation length $2\pi/\xi_1^{(cc)}$ and decay correlation length $1/\xi_0^{(cc)}$ between the bulk and slit
  systems. However, in both cases the amplitude of charge-charge correlations is smaller than the corresponding bulk value, demonstrated
  as a downwards shift on the $y$-axis. We also observe a slight phase shift between the bulk and slit system charge-charge
  correlations. The density-density correlations have a smaller decay correlation length for the short surface
  separation Figure \ref{fig:comp_correlations}(b), and larger for the wider surface separation Figure \ref{fig:comp_correlations}(c), when compared to
  the bulk values. In both cases, the amplitude is again smaller for the slit system. Figure \ref{fig:comp_correlations}(d) demonstrates the correlation
  analysis for a wide surface separation with neutral walls. We can see that the difference between
  Figure \ref{fig:comp_correlations}(c) and Figure \ref{fig:comp_correlations}(d) is minimal, demonstrating that the charge of the surfaces
  has no effect on the lateral correlation decay at the mid plane at large surface separations.  Importantly, the similarity of frequency and decay
  charge-charge correlation lengths between slit and bulk results at all surface separations (albeit with moderately large differences
  in amplitudes and a slight phase shift) demonstrate that the asymptotic behaviour lateral to the confining charged surfaces is dictated by bulk
  properties of the ionic fluid and \textit{not} by the surfaces themselves.  For completeness, the same analysis is additionally presented for
  the low concentration system in the ESI.
	\begin{figure}
		\centering
		\includegraphics[scale = 1]{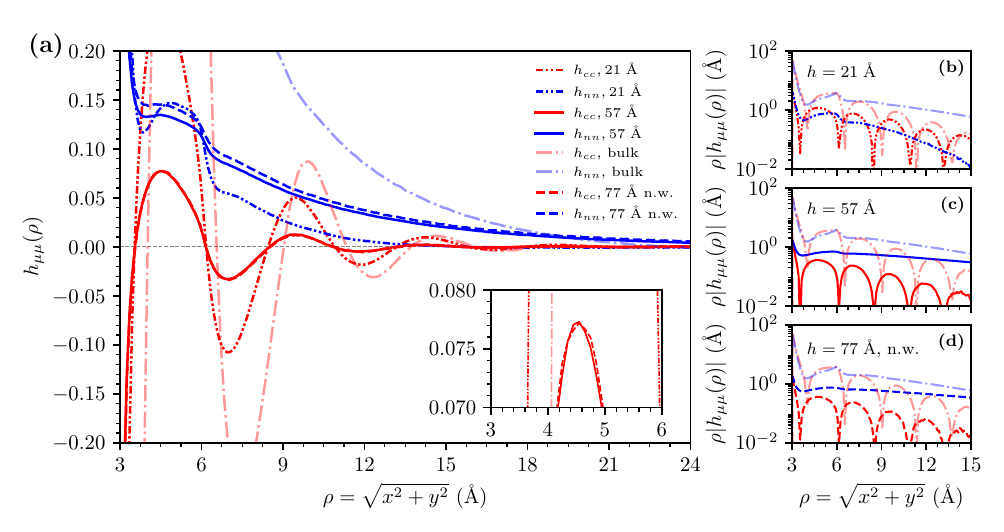}  
		\caption{{Comparisons of correlation functions obtained for the $c \approx 3.4$ M system at
		  the mid plane for $h = 21$ Å , $h = 57$ Å charged wall systems, as well as the $h = 77$ Å neutral wall (n.w.) system, with bulk
                  correlation functions. For the bulk, all correlation functions have a standard radial
                  dependence i.e. $h_{\mu\mu} \equiv f(r)$.
                  Subplots compare the bulk correlation functions with (a) all the correlation functions on a linear scale,
                  (b) $h = 21$ Å slit correlation
                  functions, (c) $h = 57$ Å slit correlation functions, and (d) $h = 77$ Å with neutral walls slit
                  correlation functions.}}
		\label{fig:comp_correlations}
	\end{figure}
}

    {
      
  \subsection{Comparison of different sPMF models at high salt concentrations}
  Here, we will compare results obtained with the sRPM($\phi_n$), and the sRPM($\phi_w$) models
  (where the latter uses the wider form of the sPMF in addition to the RPM 
  interactions).  Recall that with this latter model
  we have different combinations of attractions and repulsions between ionic species.
  To reiterate, they are denoted as, $\phi_w(a,r)$ (attractive $+-$,repulsive $++,--$), $\phi_w(a,a)$ (attractive $+-$,attractive $++,--$)
  and $\phi_w(a,0)$ (attractive $+-$, no sPMF between $++$ and $--$).
  In all cases below, we have set $d=3$ {\AA}, and used the
  ``steep'' wall description. Bulk salt concentrations were adjusted
  (via the chemical potential) to lie in the regime 1.6-2 M.   

\begin{figure}
	\centering
	\includegraphics[scale=1]{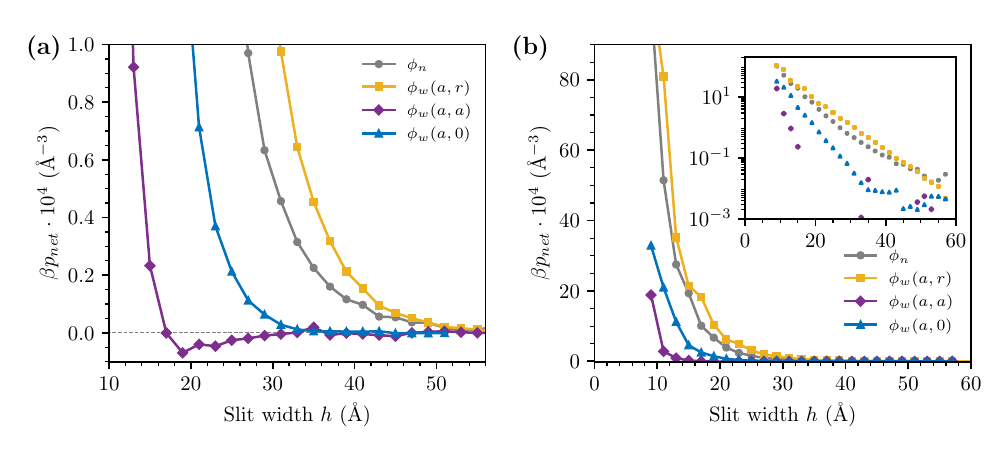}          
	\caption{{ Net pressure curves, as obtained with the sRPM($\phi_n$) (reference)
          as well as with various versions of $\phi_w$. The bulk
          salt concentrations are within the regime 1.6-2 M. Graph (a) focuses on the
          large separation regime. The negative part of the $\phi_w(a,a)$ curve has been removed in
        the inset of graph (b), that displays the net pressures on a log scale.}}
	\label{fig:sPMFcomp}
\end{figure}    
  Net pressure curves in concentrated samples (1.6-2 M), for various choices of sPMF, are summarised in
  Figure \ref{fig:sPMFcomp}. A few important observations immediately emerge:
  \begin{itemize}
  \item The long-ranged pressures that are obtained with models in which the
    additional sPMF is attractive between unlike charges, and repulsive
    between like charges, is large.
    That is, the net pressures obtained from sRPM($\phi_w(a,r)$)
    are qualitatively similar to those of sRPM($\phi_n$). The sRPM($\phi_w(a,r)$) model
    generates a somewhat stronger repulsion, but
    the difference is small.
  \item The {\em qualitative nature} of the sPMF
    is quite important. Specifically, the sRPM($\phi_w(a,a)$) model only produces a short-ranged
    repulsive regime, outside of which we even note a weak {\em attractive} interaction.
    The sRPM($\phi_w(a,0)$) model is an intermediate case, where the repulsion clearly extends
    further than with the pure RPM, but it is not as long-ranged as with $\phi_w(a,r)$
    or $\phi_n$.
  \item We do not observe any force oscillations in the investigated separation regime (irrespective of the chosen sPMF).
    SFA measurements \cite{Gebbie:13,Gebbie:15,Smith:2016}  have reported force ``jumps'', indicative of
    such oscillations. On the other hand, such
    behaviours were not observed in a recent AFM \cite{Kumar:22} study. It should be noted that possible oscillations
    due to solvent packing will be unaccounted for by our implicit solvent treatment. 
  \end{itemize}

  It is also of interest to compare the cluster tendency displayed by some of our investigated models. 
    \begin{figure}[h!]
  \centering
  \includegraphics[scale = 1]{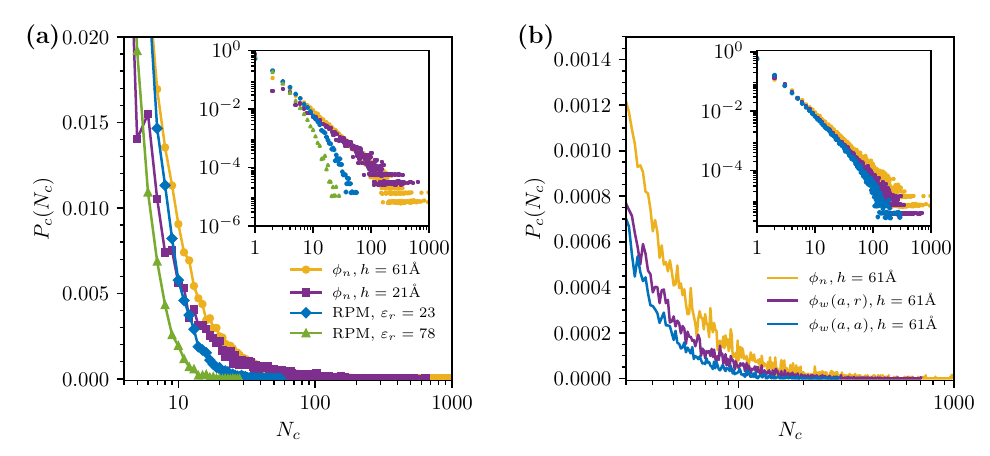}        
        \caption{{Cluster probability distributions. 
            (a) Comparing the cluster distribution ($P_c(N_c)$), at a rather wide surface
            separation (61{\AA}), of the sRPM($\phi_n$) with
            RPM:s with a uniform dielectric constant of 23 and 78.3.
            Also shown is the cluster distribution for the sRPM($\phi_n$) at
            a narrow separation, 21{\AA}. \newline
            (b) The long-ranged tail of $P_c(N_c)$, at a surface separation of 61{\AA}, for
            sRPM($\phi_n$), sRPM($\phi_w(a,r)$) and sRPM($\phi_w(a,a)$). The thick lines are
        10-point running averages.}}
	\label{fig:clustsum}
\end{figure}
  On Figure \ref{fig:clustsum} we plot cluster probability
  distributions, $P_c(N_c)$, for a range of different models. Here, $P_c$ measures
  the probability for a cluster of size $N_c$. A cluster is defined such that an ion must
  be within a distance $\delta_c$ or less, of at least one other ion within a cluster, in order to be a member of that cluster.
  We have set $\delta_c = d+2${\AA}.
  These distributions were obtained from grand canonical slit simulations, wher ethe chemical potential is adjusted
  so that the corresponding bulk solution has a concentration of about 1.8 M. The distributions are
  based on at least 50 different configurations, separated by $10^8$ attempted moves to ensure
  statistical independence. 

  We note that the RPM approach does not generate large clusters, even if the (uniform) dielectric
  constant is set as low as 23. The sRPM:s, on the other hand, are able to produce quite large ion clusters.
  Interestingly enough, this is true also for the sRPM($\phi_w(a,a)$), even though we have seen that the corresponding
  surface forces are rather short-ranged.  Nevertheless, the large cluster tail does not extend
  as far as with sRPM($\phi_w(a,r)$) or (in particular) sRPM($\phi_n$).  
  Even so, these results imply that 
  it is possibly not {\em only} the number of particles in each cluster that matters, but also the intrinsic structure
  of such clusters. This is supported by the long-ranged radial distribution tails that we observe with
  $\phi_n$ and $\phi_w(a,r)$, but not with $\phi_w(a,a)$ (see the ESI for details).
  In future work, we will scrutinise this further, making use of polymer classical
  Density Functional Theory.

  In graph (a) of Figure \ref{fig:clustsum} we also see that, as expected, very large clusters become improbable
  at short separations. It should be emphasised that this is not simply an effect of a diminishing number
  of ions within the simulation box in this regime. This is illustrated in the ESI, where we find
  a very similar cluster probability distribution at short separations, but with an increased
  lateral ($x,y$) size of the simulation box.

\section{Conclusion}

In this article we have demonstrated that a modified RPM with an added short-ranged
PMF, adjusted to provide the model more realistic saturation properties, produces surface forces
in much better agreement with experiments than can be obtained with the RPM model itself.  This 
is related to the increased prevalence of clusters, brought about by the added short-ranged solvent-induced
potential, that we, {in the initial part of this work (concentration dependence), chose to model
by a locally enhanced electrostatic coupling.
Density-density correlations appear to dominate at high salt concentrations.} 
Thus, we hypothesise that the experimentally observed anomalous screening lengths originate from 
\textit{cluster} correlations rather than \textit{simple ion} correlations.
One apparent failure of the sRPM($\phi_n$) is its inability to predict a growth in the interaction screening length 
with electrolyte concentration.  This is possibly due to the lack of many-body contributions to the ion-ion interaction, 
which potentially leads to an underestimation of the effects from concentration on particle 
interactions.  A future aim is to implement a simple many-body scheme which will provide some estimate of this
effect.

{ In the second part of this work, we established the {\em qualitative} nature of the sPMF is of crucial importance, even though
its functional form is not. Our combined surface force and bulk solution cluster analyses suggest that
the internal structure of the ion clusters may be as important as their average size. This will
be investigated more closely in future work, employing classical polymer density functional theory.}



\section*{Acknowledgements}

We thank professor Susan Perkin for sending us raw data from SFA measurements, as well as
for fruitful discussions. Professors Sture Nordholm and Christian Holm are also acknowledged for enlightening discussions.
  J.F. acknowledges financial support by the Swedish Research Council, and computational resources by the
  Lund University computer cluster organisation, LUNARC. 


\begin{mcitethebibliography}{43}
\providecommand*{\natexlab}[1]{#1}
\providecommand*{\mciteSetBstSublistMode}[1]{}
\providecommand*{\mciteSetBstMaxWidthForm}[2]{}
\providecommand*{\mciteBstWouldAddEndPuncttrue}
  {\def\EndOfBibitem{\unskip.}}
\providecommand*{\mciteBstWouldAddEndPunctfalse}
  {\let\EndOfBibitem\relax}
\providecommand*{\mciteSetBstMidEndSepPunct}[3]{}
\providecommand*{\mciteSetBstSublistLabelBeginEnd}[3]{}
\providecommand*{\EndOfBibitem}{}
\mciteSetBstSublistMode{f}
\mciteSetBstMaxWidthForm{subitem}
{(\emph{\alph{mcitesubitemcount}})}
\mciteSetBstSublistLabelBeginEnd{\mcitemaxwidthsubitemform\space}
{\relax}{\relax}

\bibitem[Israelachvili(1991)]{Israelachvili:91}
J.~N. Israelachvili, \emph{Intermolecular and Surface Forces, 2nd Ed.},
  Academic Press, London, 1991\relax
\mciteBstWouldAddEndPuncttrue
\mciteSetBstMidEndSepPunct{\mcitedefaultmidpunct}
{\mcitedefaultendpunct}{\mcitedefaultseppunct}\relax
\EndOfBibitem
\bibitem[Evans and Wennerstr{\"o}m(1994)]{Evans:94}
F.~A. Evans and H.~Wennerstr{\"o}m, \emph{The colloidal domain: where Physics,
  Chemistry, Biology and Technology meet}, VCH Publishers, New York, 1994\relax
\mciteBstWouldAddEndPuncttrue
\mciteSetBstMidEndSepPunct{\mcitedefaultmidpunct}
{\mcitedefaultendpunct}{\mcitedefaultseppunct}\relax
\EndOfBibitem
\bibitem[Holm \emph{et~al.}(2001)Holm, Kekicheff, and Podgornik]{Holm:01}
C.~Holm, P.~Kekicheff and R.~Podgornik, \emph{Electrostatic Effects in Soft
  Matter and Biophysics}, Kluwer Academic Publishers, Dordrecht, 2001\relax
\mciteBstWouldAddEndPuncttrue
\mciteSetBstMidEndSepPunct{\mcitedefaultmidpunct}
{\mcitedefaultendpunct}{\mcitedefaultseppunct}\relax
\EndOfBibitem
\bibitem[Derjaguin and Landau(1941)]{Derjaguin:41}
B.~V. Derjaguin and L.~Landau, \emph{Acta Phys. Chim. URSS}, 1941, \textbf{14},
  633--662\relax
\mciteBstWouldAddEndPuncttrue
\mciteSetBstMidEndSepPunct{\mcitedefaultmidpunct}
{\mcitedefaultendpunct}{\mcitedefaultseppunct}\relax
\EndOfBibitem
\bibitem[Verwey and Overbeek(1948)]{Verwey:48}
E.~J.~W. Verwey and J.~T.~G. Overbeek, \emph{Theory of the Stability of
  Lyophobic Colloids}, Elsevier Publishing Company Inc., Amsterdam, 1948\relax
\mciteBstWouldAddEndPuncttrue
\mciteSetBstMidEndSepPunct{\mcitedefaultmidpunct}
{\mcitedefaultendpunct}{\mcitedefaultseppunct}\relax
\EndOfBibitem
\bibitem[Attard(1993)]{Attard_1993}
P.~Attard, \emph{Phys. Rev. E}, 1993, \textbf{48}, 3604--3621\relax
\mciteBstWouldAddEndPuncttrue
\mciteSetBstMidEndSepPunct{\mcitedefaultmidpunct}
{\mcitedefaultendpunct}{\mcitedefaultseppunct}\relax
\EndOfBibitem
\bibitem[Leote~de Carvalho and Evans(1994)]{LeotedeCarvalho1994}
R.~Leote~de Carvalho and R.~Evans, \emph{Mol. Phys.}, 1994, \textbf{83},
  619–654\relax
\mciteBstWouldAddEndPuncttrue
\mciteSetBstMidEndSepPunct{\mcitedefaultmidpunct}
{\mcitedefaultendpunct}{\mcitedefaultseppunct}\relax
\EndOfBibitem
\bibitem[Forsman \emph{et~al.}(2024)Forsman, Ribar, and Woodward]{Forsman:24a}
J.~Forsman, D.~Ribar and C.~E. Woodward, \emph{Phys. Chem. Chem. Phys.}, 2024,
  DOI: 10.1039/D4CP00546E\relax
\mciteBstWouldAddEndPuncttrue
\mciteSetBstMidEndSepPunct{\mcitedefaultmidpunct}
{\mcitedefaultendpunct}{\mcitedefaultseppunct}\relax
\EndOfBibitem
\bibitem[Kirkwood(1936)]{Kirkwood1936}
J.~G. Kirkwood, \emph{Chemical Reviews}, 1936, \textbf{19}, 275–307\relax
\mciteBstWouldAddEndPuncttrue
\mciteSetBstMidEndSepPunct{\mcitedefaultmidpunct}
{\mcitedefaultendpunct}{\mcitedefaultseppunct}\relax
\EndOfBibitem
\bibitem[Torrie and Valleau(1980)]{Torrie:80}
G.~M. Torrie and J.~P. Valleau, \emph{J.~Chem. Phys.}, 1980, \textbf{73},
  5807--5816\relax
\mciteBstWouldAddEndPuncttrue
\mciteSetBstMidEndSepPunct{\mcitedefaultmidpunct}
{\mcitedefaultendpunct}{\mcitedefaultseppunct}\relax
\EndOfBibitem
\bibitem[Gebbie \emph{et~al.}(2013)Gebbie, Valtiner, Banquy, Fox, Henderson,
  and Israelachvili]{Gebbie:13}
M.~A. Gebbie, M.~Valtiner, X.~Banquy, E.~T. Fox, W.~A. Henderson and J.~N.
  Israelachvili, \emph{PNAS}, 2013, \textbf{110}, 9674--9679\relax
\mciteBstWouldAddEndPuncttrue
\mciteSetBstMidEndSepPunct{\mcitedefaultmidpunct}
{\mcitedefaultendpunct}{\mcitedefaultseppunct}\relax
\EndOfBibitem
\bibitem[Gebbie \emph{et~al.}(2015)Gebbie, Dobbs, Valtiner, and
  Israelachvili]{Gebbie:15}
M.~A. Gebbie, H.~A. Dobbs, M.~Valtiner and J.~N. Israelachvili, \emph{PNAS},
  2015, \textbf{112}, 7432--7437\relax
\mciteBstWouldAddEndPuncttrue
\mciteSetBstMidEndSepPunct{\mcitedefaultmidpunct}
{\mcitedefaultendpunct}{\mcitedefaultseppunct}\relax
\EndOfBibitem
\bibitem[Smith \emph{et~al.}(2016)Smith, Lee, and Perkin]{Smith:2016}
A.~M. Smith, A.~A. Lee and S.~Perkin, \emph{J. Phys. Chem. Lett.}, 2016,
  \textbf{7}, 2157--2163\relax
\mciteBstWouldAddEndPuncttrue
\mciteSetBstMidEndSepPunct{\mcitedefaultmidpunct}
{\mcitedefaultendpunct}{\mcitedefaultseppunct}\relax
\EndOfBibitem
\bibitem[Elliott \emph{et~al.}(2024)Elliott, Gregory, Robertson, Craig, Webber,
  Wanless, and Page]{Elliot:24}
G.~R. Elliott, K.~P. Gregory, H.~Robertson, V.~S. Craig, G.~B. Webber, E.~J.
  Wanless and A.~J. Page, \emph{Chemical Physics Letters}, 2024, \textbf{843},
  141190\relax
\mciteBstWouldAddEndPuncttrue
\mciteSetBstMidEndSepPunct{\mcitedefaultmidpunct}
{\mcitedefaultendpunct}{\mcitedefaultseppunct}\relax
\EndOfBibitem
\bibitem[Coupette \emph{et~al.}(2018)Coupette, Lee, and Härtel]{coupette:18}
F.~Coupette, A.~A. Lee and A.~Härtel, \emph{Phys. Rev. Lett.}, 2018,
  \textbf{121}, 075501\relax
\mciteBstWouldAddEndPuncttrue
\mciteSetBstMidEndSepPunct{\mcitedefaultmidpunct}
{\mcitedefaultendpunct}{\mcitedefaultseppunct}\relax
\EndOfBibitem
\bibitem[Rotenberg \emph{et~al.}(2018)Rotenberg, Bernard, and
  Hansen]{Rotenberg:18}
B.~Rotenberg, O.~Bernard and J.-P. Hansen, \emph{J. Phys. Condens. Matter},
  2018, \textbf{30}, 054005\relax
\mciteBstWouldAddEndPuncttrue
\mciteSetBstMidEndSepPunct{\mcitedefaultmidpunct}
{\mcitedefaultendpunct}{\mcitedefaultseppunct}\relax
\EndOfBibitem
\bibitem[Kjellander(2018)]{Kjellander2018}
R.~Kjellander, \emph{J. Chem. Phys.}, 2018, \textbf{148}, 193701\relax
\mciteBstWouldAddEndPuncttrue
\mciteSetBstMidEndSepPunct{\mcitedefaultmidpunct}
{\mcitedefaultendpunct}{\mcitedefaultseppunct}\relax
\EndOfBibitem
\bibitem[Coles \emph{et~al.}(2020)Coles, Park, Nikam, Kanduc, Dzubiella, and
  Rotenberg]{Coles:20}
S.~W. Coles, C.~Park, R.~Nikam, M.~Kanduc, J.~Dzubiella and B.~Rotenberg,
  \emph{J. Phys. Chem. B}, 2020, \textbf{124}, 1778--1786\relax
\mciteBstWouldAddEndPuncttrue
\mciteSetBstMidEndSepPunct{\mcitedefaultmidpunct}
{\mcitedefaultendpunct}{\mcitedefaultseppunct}\relax
\EndOfBibitem
\bibitem[Zeman \emph{et~al.}(2020)Zeman, Kondrat, and Holm]{Zeman:20}
J.~Zeman, S.~Kondrat and C.~Holm, \emph{Chem. Commun.}, 2020, \textbf{56},
  15635--15638\relax
\mciteBstWouldAddEndPuncttrue
\mciteSetBstMidEndSepPunct{\mcitedefaultmidpunct}
{\mcitedefaultendpunct}{\mcitedefaultseppunct}\relax
\EndOfBibitem
\bibitem[Cats \emph{et~al.}(2021)Cats, Evans, H{\"a}rtel, and van
  Roij]{Cats:21}
P.~Cats, R.~Evans, A.~H{\"a}rtel and R.~van Roij, \emph{J. Chem. Phys.}, 2021,
  \textbf{154}, 124504\relax
\mciteBstWouldAddEndPuncttrue
\mciteSetBstMidEndSepPunct{\mcitedefaultmidpunct}
{\mcitedefaultendpunct}{\mcitedefaultseppunct}\relax
\EndOfBibitem
\bibitem[H{\"a}rtel \emph{et~al.}(2023)H{\"a}rtel, B\"ultmann, and
  Coupette]{Hartel:23}
A.~H{\"a}rtel, M.~B\"ultmann and F.~Coupette, \emph{Phys. Rev. Lett.}, 2023,
  \textbf{130}, 108202\relax
\mciteBstWouldAddEndPuncttrue
\mciteSetBstMidEndSepPunct{\mcitedefaultmidpunct}
{\mcitedefaultendpunct}{\mcitedefaultseppunct}\relax
\EndOfBibitem
\bibitem[Gaddam and Ducker(2019)]{Gaddam:19}
P.~Gaddam and W.~Ducker, \emph{Langmuir}, 2019, \textbf{35}, 5719--5727\relax
\mciteBstWouldAddEndPuncttrue
\mciteSetBstMidEndSepPunct{\mcitedefaultmidpunct}
{\mcitedefaultendpunct}{\mcitedefaultseppunct}\relax
\EndOfBibitem
\bibitem[Yuan \emph{et~al.}(2022)Yuan, Deng, Zhu, Liu, and Craig]{Yuan:22}
H.~Yuan, W.~Deng, X.~Zhu, G.~Liu and V.~S.~J. Craig, \emph{Langmuir}, 2022,
  \textbf{38}, 6164--6173\relax
\mciteBstWouldAddEndPuncttrue
\mciteSetBstMidEndSepPunct{\mcitedefaultmidpunct}
{\mcitedefaultendpunct}{\mcitedefaultseppunct}\relax
\EndOfBibitem
\bibitem[Reinertsen \emph{et~al.}(2024)Reinertsen, Kewalramani,
  Jiménez-Ángeles, Weigand, Bedzyk, and Olvera de~la Cruz]{Reinertsen2024}
R.~J.~E. Reinertsen, S.~Kewalramani, F.~Jiménez-Ángeles, S.~J. Weigand, M.~J.
  Bedzyk and M.~Olvera de~la Cruz, \emph{Proceedings of the National Academy of
  Sciences}, 2024, \textbf{121}, year\relax
\mciteBstWouldAddEndPuncttrue
\mciteSetBstMidEndSepPunct{\mcitedefaultmidpunct}
{\mcitedefaultendpunct}{\mcitedefaultseppunct}\relax
\EndOfBibitem
\bibitem[Robertson \emph{et~al.}(2023)Robertson, Elliott, Nelson, Le~Brun,
  Webber, Prescott, Craig, Wanless, and Willott]{Robertson2023}
H.~Robertson, G.~R. Elliott, A.~R.~J. Nelson, A.~P. Le~Brun, G.~B. Webber,
  S.~W. Prescott, V.~S.~J. Craig, E.~J. Wanless and J.~D. Willott,
  \emph{Physical Chemistry Chemical Physics}, 2023, \textbf{25},
  24770–24782\relax
\mciteBstWouldAddEndPuncttrue
\mciteSetBstMidEndSepPunct{\mcitedefaultmidpunct}
{\mcitedefaultendpunct}{\mcitedefaultseppunct}\relax
\EndOfBibitem
\bibitem[Kumar \emph{et~al.}(2022)Kumar, Cats, Alotaibi, Ayirala, Yousef, {van
  Roij}, Siretanu, and Mugele]{Kumar:22}
S.~Kumar, P.~Cats, M.~B. Alotaibi, S.~C. Ayirala, A.~A. Yousef, R.~{van Roij},
  I.~Siretanu and F.~Mugele, \emph{J. Colloid Interface Sci.}, 2022,
  \textbf{622}, 819--827\relax
\mciteBstWouldAddEndPuncttrue
\mciteSetBstMidEndSepPunct{\mcitedefaultmidpunct}
{\mcitedefaultendpunct}{\mcitedefaultseppunct}\relax
\EndOfBibitem
\bibitem[Ma \emph{et~al.}(2015)Ma, Forsman, and Woodward]{Ma:15}
K.~Ma, J.~Forsman and C.~E. Woodward, \emph{The Journal of Chemical Physics},
  2015, \textbf{142}, 174704\relax
\mciteBstWouldAddEndPuncttrue
\mciteSetBstMidEndSepPunct{\mcitedefaultmidpunct}
{\mcitedefaultendpunct}{\mcitedefaultseppunct}\relax
\EndOfBibitem
\bibitem[Hasted \emph{et~al.}(2004)Hasted, Ritson, and
  Collie]{hasted_dielectric_2004}
J.~B. Hasted, D.~M. Ritson and C.~H. Collie, \emph{J. Chem. Phys.}, 2004,
  \textbf{16}, 1--21\relax
\mciteBstWouldAddEndPuncttrue
\mciteSetBstMidEndSepPunct{\mcitedefaultmidpunct}
{\mcitedefaultendpunct}{\mcitedefaultseppunct}\relax
\EndOfBibitem
\bibitem[de~Souza \emph{et~al.}(2022)de~Souza, Kornyshev, and
  Bazant]{deSouza2022}
J.~de~Souza, A.~A. Kornyshev and M.~Z. Bazant, \emph{J. Chem. Phys.}, 2022,
  \textbf{156}, year\relax
\mciteBstWouldAddEndPuncttrue
\mciteSetBstMidEndSepPunct{\mcitedefaultmidpunct}
{\mcitedefaultendpunct}{\mcitedefaultseppunct}\relax
\EndOfBibitem
\bibitem[Conway and Marshall(1983)]{Conway83}
B.~Conway and S.~Marshall, \emph{Aust. J. Chem.}, 1983, \textbf{36},
  2145--2161\relax
\mciteBstWouldAddEndPuncttrue
\mciteSetBstMidEndSepPunct{\mcitedefaultmidpunct}
{\mcitedefaultendpunct}{\mcitedefaultseppunct}\relax
\EndOfBibitem
\bibitem[Bonthuis \emph{et~al.}(2012)Bonthuis, Gekle, and Netz]{Bonthuis2012}
D.~J. Bonthuis, S.~Gekle and R.~R. Netz, \emph{Langmuir}, 2012, \textbf{28},
  7679–7694\relax
\mciteBstWouldAddEndPuncttrue
\mciteSetBstMidEndSepPunct{\mcitedefaultmidpunct}
{\mcitedefaultendpunct}{\mcitedefaultseppunct}\relax
\EndOfBibitem
\bibitem[Danielewicz-Ferchmin \emph{et~al.}(2013)Danielewicz-Ferchmin,
  Banachowicz, and Ferchmin]{DanielewiczFerchmin2013}
I.~Danielewicz-Ferchmin, E.~Banachowicz and A.~Ferchmin, \emph{J. Mol. Liq.},
  2013, \textbf{187}, 157–164\relax
\mciteBstWouldAddEndPuncttrue
\mciteSetBstMidEndSepPunct{\mcitedefaultmidpunct}
{\mcitedefaultendpunct}{\mcitedefaultseppunct}\relax
\EndOfBibitem
\bibitem[Adar \emph{et~al.}(2018)Adar, Markovich, Levy, Orland, and
  Andelman]{adar_dielectric_2018}
R.~M. Adar, T.~Markovich, A.~Levy, H.~Orland and D.~Andelman, \emph{J. Chem.
  Phys.}, 2018, \textbf{149}, 054504\relax
\mciteBstWouldAddEndPuncttrue
\mciteSetBstMidEndSepPunct{\mcitedefaultmidpunct}
{\mcitedefaultendpunct}{\mcitedefaultseppunct}\relax
\EndOfBibitem
\bibitem[Underwood and Bourg(2022)]{Underwood2022}
T.~R. Underwood and I.~C. Bourg, \emph{The Journal of Physical Chemistry B},
  2022, \textbf{126}, 2688--2698\relax
\mciteBstWouldAddEndPuncttrue
\mciteSetBstMidEndSepPunct{\mcitedefaultmidpunct}
{\mcitedefaultendpunct}{\mcitedefaultseppunct}\relax
\EndOfBibitem
\bibitem[Ribar \emph{et~al.}(2024)Ribar, Woodward, Nordholm, and
  Forsman]{Ribar:24}
D.~Ribar, C.~E. Woodward, S.~Nordholm and J.~Forsman, \emph{J. Phys.Chem.
  Lett.}, 2024, \textbf{15}, 8326–8333\relax
\mciteBstWouldAddEndPuncttrue
\mciteSetBstMidEndSepPunct{\mcitedefaultmidpunct}
{\mcitedefaultendpunct}{\mcitedefaultseppunct}\relax
\EndOfBibitem
\bibitem[Stenberg and Forsman(2021)]{Stenberg:2021a}
S.~Stenberg and J.~Forsman, \emph{Langmuir}, 2021, \textbf{37},
  14360--14368\relax
\mciteBstWouldAddEndPuncttrue
\mciteSetBstMidEndSepPunct{\mcitedefaultmidpunct}
{\mcitedefaultendpunct}{\mcitedefaultseppunct}\relax
\EndOfBibitem
\bibitem[Crothers \emph{et~al.}(2021)Crothers, Li, and Radke]{Crothers:21}
A.~R. Crothers, C.~Li and C.~Radke, \emph{Advances in Colloid and Interface
  Science}, 2021, \textbf{288}, 102335\relax
\mciteBstWouldAddEndPuncttrue
\mciteSetBstMidEndSepPunct{\mcitedefaultmidpunct}
{\mcitedefaultendpunct}{\mcitedefaultseppunct}\relax
\EndOfBibitem
\bibitem[Evans \emph{et~al.}(1994)Evans, Leote~de Carvalho, Henderson, and
  Hoyle]{Evans1994}
R.~Evans, R.~J.~F. Leote~de Carvalho, J.~R. Henderson and D.~C. Hoyle,
  \emph{The Journal of Chemical Physics}, 1994, \textbf{100}, 591603\relax
\mciteBstWouldAddEndPuncttrue
\mciteSetBstMidEndSepPunct{\mcitedefaultmidpunct}
{\mcitedefaultendpunct}{\mcitedefaultseppunct}\relax
\EndOfBibitem
\bibitem[Degreve and da~Silva(1999)]{Degreve:99}
L.~Degreve and F.~L.~B. da~Silva, \emph{The Journal of Chemical Physics}, 1999,
  \textbf{111}, 5150--5156\relax
\mciteBstWouldAddEndPuncttrue
\mciteSetBstMidEndSepPunct{\mcitedefaultmidpunct}
{\mcitedefaultendpunct}{\mcitedefaultseppunct}\relax
\EndOfBibitem
\bibitem[Komori and Terao(2023)]{Komori:23}
K.~Komori and T.~Terao, \emph{Chemical Physics Letters}, 2023, \textbf{825},
  140627\relax
\mciteBstWouldAddEndPuncttrue
\mciteSetBstMidEndSepPunct{\mcitedefaultmidpunct}
{\mcitedefaultendpunct}{\mcitedefaultseppunct}\relax
\EndOfBibitem
\bibitem[Auffinger \emph{et~al.}(2007)Auffinger, Cheatham, and
  Vaiana]{Auffinger:07}
P.~Auffinger, T.~E. Cheatham and A.~C. Vaiana, \emph{Journal of Chemical Theory
  and Computation}, 2007, \textbf{3}, 1851--1859\relax
\mciteBstWouldAddEndPuncttrue
\mciteSetBstMidEndSepPunct{\mcitedefaultmidpunct}
{\mcitedefaultendpunct}{\mcitedefaultseppunct}\relax
\EndOfBibitem
\bibitem[Tong \emph{et~al.}(2023)Tong, Peng, Kontogeorgis, and Liang]{Tong:23}
J.~Tong, B.~Peng, G.~M. Kontogeorgis and X.~Liang, \emph{Journal of Molecular
  Liquids}, 2023, \textbf{371}, 121086\relax
\mciteBstWouldAddEndPuncttrue
\mciteSetBstMidEndSepPunct{\mcitedefaultmidpunct}
{\mcitedefaultendpunct}{\mcitedefaultseppunct}\relax
\EndOfBibitem
\end{mcitethebibliography}
\providecommand*{\mcitethebibliography}{\thebibliography}
\csname @ifundefined\endcsname{endmcitethebibliography}
{\let\endmcitethebibliography\endthebibliography}{}

\includepdf[pages=-]{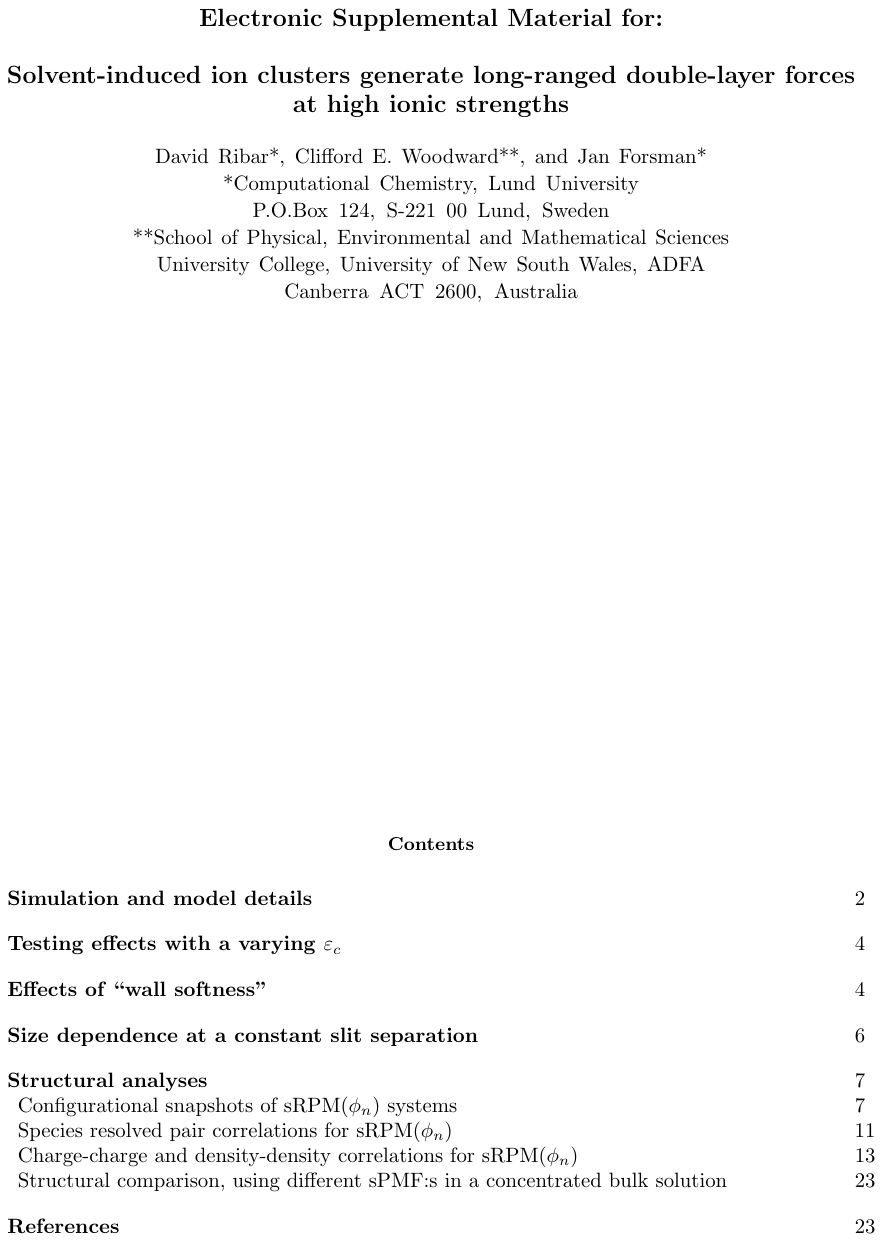}

\end{document}